\documentclass[twocolumn,prb]{revtex4-1}

\usepackage{eurosym,amsmath,amssymb}
\usepackage{bm,epsfig,graphicx,epstopdf}
\usepackage[usenames,dvipsnames]{xcolor}
\usepackage[colorlinks]{hyperref}
\usepackage[normalem]{ulem}

\hypersetup{colorlinks,citecolor=blue,linkcolor=blue,urlcolor=blue}

\begin{document}

\title{Energy levels repulsion by spin-orbit coupling in two-dimensional Rydberg excitons}
\author{V. A. Stephanovich}
\affiliation{Institute of Physics, Opole University, Opole, 45-052, Poland}
\author{E. Ya. Sherman}
\affiliation{Department of Physical Chemistry, University of the Basque Country UPV/EHU,
48080 Bilbao, Spain}
\affiliation{IKERBASQUE Basque Foundation for Science, Bilbao, Spain}
\author{N. T. Zinner}
\affiliation{Aarhus Institute of Advanced Studies, Aarhus University, DK-8000 Aarhus C, Denmark}
\affiliation{Department of Physics and Astronomy, Aarhus University, DK-8000 Aarhus C, Denmark}
\author{O. V. Marchukov}
\affiliation{School of Electrical Engineering, Faculty of Engineering, Tel Aviv University, 6997801, Tel Aviv, Israel}

\date{\today}

\begin{abstract}
We study the effects of Rashba spin-orbit coupling  on two-dimensional Rydberg exciton systems.
Using analytical and numerical arguments we demonstrate that this coupling considerably 
modifies the wave functions and leads to a level repulsion that results in a deviation from the Poissonian
statistics of the adjacent level distance distribution. This signifies the crossover to 
non-integrability of the system and hints on the possibility of quantum chaos emerging. 
Such a behavior strongly differs  from the classical realization, where spin-orbit coupling produces highly 
entangled, chaotic electron trajectories in an exciton.
We also calculate the oscillator strengths and show that randomization appears in the
transitions between states with different total momenta. 
\end{abstract}

\maketitle

\section{Introduction}

Recent discovery of quantum chaos in the spectra of Rydberg excitons  
\cite{Assmann2016, Ostrovskaya2016} posed a question about the qualitative 
role of relatively weak effects beyond a simple Coulomb interaction in exciton physics.
It has been demonstrated~\cite{Assmann2016, Schweiner2017, Schweiner2017a} that the Rydberg 
excitons in Cu$_{2}$O crystal subjected to an external magnetic field breaks all antiunitary 
symmetries and may show different types of statistics of the distances between the adjacent 
levels in the energy levels spectra. The latter is a well-known signature of 
quantum chaos~\cite{Gutzwiller, Reichl, Haake, Stockmann}, usually
expected at quantization of a classical dynamical system exhibiting a chaotic behavior. 
Experimental observation of salient
features of quantum chaos in other systems such as cold atomic gases 
\cite{frisch2014} and exciton-polariton billiards \cite{Gao2015} strongly stimulates 
the interest in understanding new features in their semiclassical realizations, 
including the appearance of chaos. We note that 
whereas chaotic behavior of many classical systems is well understood, 
its quantum mapping remains intricate.\cite{Gutzwiller,Reichl,Haake,Stockmann} 

From this point of view, the spin-orbit coupling (SOC) 
is of special interest since, as
we demonstrate here, it strongly influences the spectral statistics
in the highly excited states of two-dimensional (2D) excitons. In particular, it was proposed that 
it can engender the chaotic behavior by lowering the symmetry of the initial system.\cite{Ostrovskaya2016} 
This is particularly important as the SOC in excitons generates a great variety 
of phenomena \cite{Dyakonov08,Durnev,Vishnevsky,High} and
a detailed analysis of the SOC effects in the spectra of low-energy states was
presented in Ref. [\onlinecite{Grimaldi}]. {We mention also one more interesting physical 
mechanism acting, in some aspects, similarly to the SOC on the 2D excitons properties. 
This mechanism uses Berry curvature and lifts the time-reversal symmetry yielding 
the splitting of the energy levels 
with opposite angular momenta.\cite{zhou15} Below we briefly compare the impact 
of the two above mechanisms on the Rydberg states of our interest.}

Here we study how SOC qualitatively modifies the Rydberg states,
observe qualitative changes in the level statistics, and discuss their relation to
the possible chaos features dependent on the system symmetry and, accordingly, to 
the number of integrals of motion.  

Without SOC, all quantum properties of two-dimensional hydrogen-like systems are
well-established, also in the relativistic domain.\cite{Portnoi} 
In non-relativistic realizations, the system possesses three integrals
of motion: the energy, the angular momentum, and the Runge-Lenz vector. They fully
define the dynamics, yield the high energy levels degeneracy and assure
system stability against chaotic behavior.
Here we focus on the interplay between the spin degree of freedom and the orbital 
motion that may lead to breaking of the system integrability resulting from 
the spin back action on the orbital motion. Note that this spin back action, 
being considered quasiclassically, generates chaotic electron trajectories 
in 2D exciton with SOC.\cite{mypccp2018}

\begin{figure}[t]
\begin{center}
\includegraphics*[width=0.8\columnwidth]{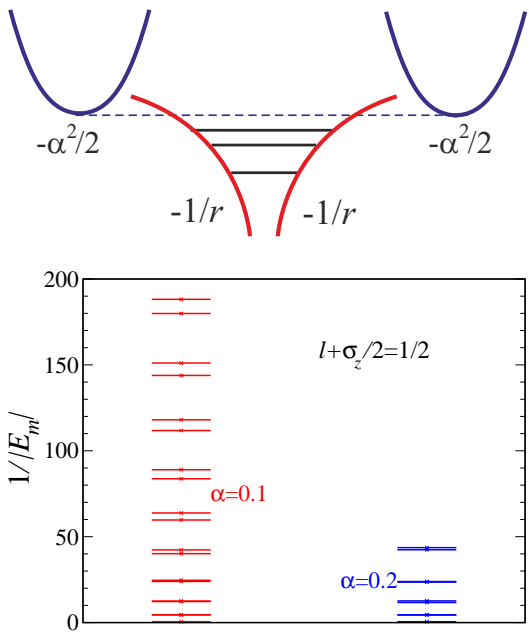}\\
\vspace{0.4cm}
\includegraphics*[width=0.8\columnwidth]{figure2}
\end{center}
\caption{Upper panel: Schematic plot of the discrete spectrum and 
Rashba-spectrum parabolas.  
The transition to delocalized states occurs 
at energies higher than $-\alpha^{2}/2.$
Lower panel: Bound states spectra $E_m$ with $l+\sigma_{z}/2=1/2$ at $\alpha=0.1$
and $\alpha=0.2$, demonstrating the absence of the degeneracy with
the increase in $m$. }
\label{fig:spectrum}
\end{figure}

\section{General formalism}
The Hamiltonian describing 2D exciton can be presented in the form 
${\mathcal H}={\mathcal H}_0+{\mathcal H}_{\mathrm{so}}$. 
The spin-independent part reads
\begin{equation}
{\mathcal H}_0=\frac{p^{2}}{2\mu }-\frac{e^{2}}{\kappa r},  \label{ham_{1}}
\end{equation}%
where $\mathbf{r}=(x,y)$ is the electron position, $\mathbf{p}=-i\hbar {\bm\nabla}$ 
is the momentum, $\mu $ is its
mass, $e$ is the charge, and $\kappa $ is the dielectric
constant of a host crystal. The SOC has the Rashba form [\onlinecite{Bychkov}],
arising in 2D structures as a result of the spatial structural inversion asymmetry:\cite{Bychkov,Fabian} 
\begin{equation}
{\mathcal H}_{\mathrm{so}}=
\frac{\alpha }{\hbar }\left( p_{x}{\sigma }_{y}-p_{y}{\sigma}_{x}\right),  
\label{hso}
\end{equation}%
with $\alpha$ being the coupling constant and $\sigma _{i}$ ($i=x,y,z$)
being the corresponding Pauli matrices. 

In the presence of spin-orbit coupling \eqref{hso}, the integral of motion corresponding to
the total angular momentum becomes $l+{\sigma}_z/2,$ where $l$ 
is the $z-$component of the orbital angular momentum. The
Runge-Lenz vector, having the 
form $\mathbf{r}/r-\left( \left[ \mathbf{p},\mathbf{l}\right]-\left[ \mathbf{l},\mathbf{p}\right]\right)/2$ 
in the SOC absence, does not have a conserved counterpart here. The discrete energy spectrum
ends at the bottom of the conduction band $-\mu\alpha ^{2}/2\hbar ^{2}.$ Nevertheless, 
the number of the localized states is still infinite, see Ref. [\onlinecite{Chaplik}] where the case of 
a weak attractive potential has been considered. 
The Hamiltonian \eqref{hso} brings a
new feature such as the anomalous spin-dependent velocity \cite{Adams} 
$\mathbf{v}_{\mathrm{so}%
}\equiv i[H_{\mathrm{so}},\mathbf{r}]/\hbar =\alpha \left( {\sigma }_{y},-{%
\sigma }_{x}\right) /\hbar $. Two other SOC characteristics of
interest \cite{Fabian} are the spin precession with the rate $2p\alpha /\hbar ^{2}$
and the corresponding length $l_{\mathrm{so}}=\hbar ^{2}/\mu\alpha $,
necessary for electron to essentially rotate the spin.  Below we
set units as $e^{2}/\kappa=\mu=\hbar=1$.

We are seeking the wave function of 2D exciton with SOC in the form of the 
infinite series (cf. [\onlinecite{Grimaldi}]) over complete set of discrete eigenstates 
of Hamiltonian ${\mathcal H}_0$ [\onlinecite{Yang}]:
\begin{eqnarray}
&&{\bm\psi}_{l}^{[m]}\left( r,\varphi \right) =e^{il\varphi }\sum_{n=l+1}%
\left[ 
\begin{array}{c}
c_{l,n\uparrow}^{[m]}R_{n,l}(r) \\ 
0
\end{array}%
\right] +  \nonumber \\
&&e^{i(l+1)\varphi }\sum_{n=l+2}\left[ 
\begin{array}{c}
0 \\ 
c_{l,n\downarrow }^{[m]}R_{n,l+1}(r)
\end{array}%
\right].
\label{psispinor}
\end{eqnarray}%

Here we take into account the conservation of total angular momentum $l+{\sigma}_{z}/2$. 
Index $m$ enumerates the eigenstates for given $l$ in the ascending energy order. Radial functions 
$R_{n,l}(r)$ correspond to the eigenstates of the
Hamiltonian ${\mathcal H}_0$ and can be expressed as: \cite{Yang}
\begin{eqnarray}
R_{n,l}\left( r\right) &=&N_{n,l}r_{n}^{\left\vert l\right\vert
}e^{-r_{n}/2}\,_{1}F_{1}\left( -n+\left\vert l\right\vert +1,2\left\vert
l\right\vert +1,r_{n}\right)  \nonumber \\
N_{n,l} &=&\frac{\beta _{n}}{\left( 2\left\vert l\right\vert \right) !}\left[ 
\frac{\left( n+\left\vert l\right\vert -1\right) !}{(2n-1)\left(
n-\left\vert l\right\vert -1\right) !}\right] ^{1/2},
\end{eqnarray}%
where dimensionless $\beta_{n}={2}/\left( n-1/2\right),$ $r_{n}=r\beta _{n}$ and $%
l=-(n-1),...,n-1.$ The expansion coefficients are the
eigenvectors of the Hamiltonian: 
\begin{equation}
\left[ 
\begin{array}{cc}
\mathcal{H}_{l}^{\uparrow \uparrow } & \mathcal{H}_{l}^{\uparrow \downarrow }
\\ 
\mathcal{H}_{l}^{\downarrow \uparrow } & \mathcal{H}_{l}^{\downarrow
\downarrow }%
\end{array}%
\right] \left[ 
\begin{array}{c}
\mathbf{c}_{l,n\uparrow}^{[m]} \\ 
\mathbf{c}_{l,n\downarrow}^{[m]}
\end{array}%
\right] =E_{m}\left[ 
\begin{array}{c}
\mathbf{c}_{l,n\uparrow}^{[m]} \\ 
\mathbf{c}_{l,n\downarrow}^{[m]}
\end{array}%
\right].
\end{equation}%
The blocks $\mathcal{H}^{\uparrow \uparrow }$ and $\mathcal{H}%
^{\downarrow \downarrow }$ are diagonal with the elements given by the
corresponding eigenenergies $E_{n}=-1/2(n-1/2)^2$ for 
2D exciton without SOC. The blocks $\mathcal{H}^{\uparrow \downarrow }$ and 
$\mathcal{H}^{\downarrow \uparrow }$ couple spin-up and spin-down states due
to the Rashba interaction. The off-diagonal elements are given by:
\begin{eqnarray}
&&\mathcal{H}_{l,n_{1},n_{2}}^{\uparrow \downarrow }=\alpha \times \label{offdiag} \\
&&\int R_{n_{1},l}( r )\left( \frac{d}{dr}R_{n_{2},l+1}(r
)+\frac{l+1}{r}R_{n_{2},l+1}(r)\right) rdr. 
\nonumber
\end{eqnarray}%
Note that the off-diagonal elements \eqref{offdiag} are usually small since $R_{nl}$
are rapidly oscillating functions.

\section{Analytical results with semiclassical approach}
Although the matrix elements \eqref{offdiag} can be evaluated analytically using 
properties of hypergeometric functions,\cite{land3,Gradshteyn} this calculation is quite cumbersome.
Thus, it is instructive to apply the simpler 
semiclassical approach for this purpose. This approach is valid at large-$l$, which is of interest 
here since for the Rydberg states with $n\gg 1$, typical values of angular momentum are also 
large $l\gg 1$.
Semiclassical evaluation of matrix elements \eqref{offdiag} 
is also useful as it permits to trace analytically the main matrix elements leading to 
the level repulsion. After separation of angular and radial 
variables, the effective Coulomb potential can be rendered as  
$\widetilde{U}(r)=-1/r+l^{2}/2r^{2},$ and at a given energy $E$, 
the semiclassical return points with $\widetilde{U}(r)=E$ are given by: 
\begin{equation}
r=\frac{l^{2}}{1\pm \sqrt{1+2l^{2}E}}.
\end{equation}%
For $l^{2}|E|\ll 1,$ the points are $r_{\min}=l^{2}/2 $ and $r_{\max }=-1/E,$ 
respectively. Forces acting on the
electron at the return points are $F\left( r_{\min }\right) =-4/l^{4}$ and $%
F\left( r_{\max }\right) =E^{2}\ll F\left( r_{\min}\right).$ As a result,
the non-semiclassical domain near $r_{\max }$ is determined by: 
$r_{\max}-r \le E^{-2/3},$ confirming validity of the semiclassical 
approximation since $r_{\max}\sim 1/E.$

The minimum of $\widetilde{U}(r)$ is achieved at $r_{l}=l^{2}$ with $r_{l+1}-r_{l}=2l+1,$
and $\widetilde{U}(r_{l})=-1/2l^{2}.$ The near-the-minimum oscillation frequencies are
given by $\Omega =\sqrt{\widetilde{U}^{\prime\prime}(r_{l})}=1/l^{3}.$ The corresponding oscillator
length $\ell =\sqrt{1/\Omega}=l^{3/2}\gg 2l$ implies that the states with orbital momenta $l$ and $l+1$ have
a large spatial overlap. Taking into account that $\Omega \ll \left|\widetilde{U}(r_{l})\right|,$ we
conclude that these are well-described in terms of harmonic oscillators 
centered at the points $r_{l}$ or $r_{l+1}$. In this approximation we obtain%
\begin{equation}
R_{n,l}(r)=\sqrt{\frac{1}{2\pi r_{l}}}\phi_{n^{[r]}}(y),
\label{Rnlsemicl}
\end{equation}%
where $y=\left(r-r_{l}\right)/l^{3/2}$ and $\phi_{n^{[r]}}(y)$ 
is the oscillator wavefunction with the radial quantum number $n^{[r]}=n-l-1$.
With these functions one can evaluate the integrals in Eq. (\ref{offdiag})
in semiclassical approach.  We begin with corresponding overlap of shifted wavefunction:
\begin{eqnarray}
&&\int \phi _{n^{[r]}_{1}}(y)\phi _{n^{[r]}_{2}}(y)dy= \\
&&N_{1}N_{2}e^{-\widetilde{l}^{2}}\int
e^{-y^{2}}H_{n^{[r]}_{1}}\left( y+\widetilde{l}\right) H_{n^{[r]}_{2}}\left( y-%
\widetilde{l}\right) dy \nonumber 
\end{eqnarray}
with normalization coefficients $N_{1}$ and $N_{2}$ and $\widetilde{l}%
=1/l^{1/2},$ with $H_{n^{[r]}_{1}}\left( y+\widetilde{l}\right) $ and $%
H_{n^{[r]}_{2}}\left( y-\widetilde{l}\right) $ being the Hermitian polynomials.\cite{land3, Gradshteyn}
 
The integration yields \cite{Gradshteyn}
\begin{eqnarray}
&&\int e^{-y^{2}}H_{j}\left(y+z_{1}\right) H_{k}\left(y+z_{2}\right)dy= \label{GRint}\\ 
&& 2^{k}\sqrt{\pi}j!z_{2}^{k-j}L_{j}^{k-j}\left(-2z_{1}z_{2}\right), \nonumber
\end{eqnarray}
where $L_{j}^{k-j}\left(-2z_{1}z_{2}\right)$ are Laguerre polynomials.\cite{land3, Gradshteyn}
In our case $\widetilde{l}^{2}\ll 1$, we can use the limit
$L_{j}^{k-j}\left(0\right) =k!/j!\left(k-j\right)!$ to obtain:
\begin{equation}
\int \phi _{n^{[r]}_{1}}(x)\phi _{n^{[r]}_{2}}(x)dx=
\sqrt{\frac{2^{n^{[r]}_{2}}{n^{[r]}_{2}!}}{2^{n^{[r]}_{1}}n^{[r]}_{1}!}}
\frac{\left(-\widetilde{l}\right)^{n^{[r]}_{2}-n^{[r]}_{1}}}
{\sqrt{\left( n^{[r]}_{2}-n^{[r]}_{1}\right)!}}.
\end{equation}
Integrals of the type 
\begin{equation}
\int R_{n_{1},l}( r )\frac{d}{dr}R_{n_{2},l+1}(r)d^{2} r
\end{equation}
can be calculated with Eqs. (\ref{Rnlsemicl}) and (\ref{GRint}) using the identity
\begin{equation}
\frac{d}{dy}\phi_{n^{[r]}}(y)=\frac{1}{\sqrt{2}}
\left[\sqrt{n^{[r]}}\phi_{n^{[r]}-1}(y)-\sqrt{n^{[r]}+1}\phi_{n^{[r]}+1}(y)\right].
\end{equation}
As a result, the $l-$dependence of the SOC matrix element is given by $\sim l^{-\left(n^{[r]}_{2}-n^{[r]}_{1}\right)/2-1}$. 
Taking into account that the oscillator frequency is $\Omega=l^{-3}$, we obtain that two states are strongly 
coupled, that is $\left|\mathcal{H}_{l,n_{1},n_{2}}^{\uparrow \downarrow}\right|/\Omega>1$ 
if $l^{2-\left(n^{[r]}_{2}-n^{[r]}_{1}\right)/2}\ge 1/\alpha$. As a result, statistical distribution 
of the levels in the corresponding domain is noticeably modified by the spin-orbit coupling. 

Within the above semiclassical approach, the strong SOC conditions can be
reformulated as $r> 1/\alpha$ for the considerable spin rotation and  $v\sim \alpha $ 
for the effect of spin-orbit coupling on the electron velocity. {Taking into account
that $r_l\sim l^{2}$ and $v\sim 1/l$, the second relation can be rewritten as $l >1/\alpha$.}
This condition coincides with $|\widetilde{U}(r_{l})|<\alpha^{2}.$

\section{Numerical results: non-Poissonian spectral statistics}

We begin with characterizing the system 
spectrum. As shown in Fig. \ref{fig:spectrum}(lower panel), the infinite degeneracy
typical for 2D $U(r)=-1/r$ potential disappears due to the SOC presence, meaning that we observe the level 
repulsion. Here the state with $m=l+1$ remains nondegenerate while other
states produce split doublets with the splitting proportional to $\alpha^2$
for the first doublet (since $\mathcal{H}_{l,l+2,l+3}^{\uparrow \downarrow}=0)$   
or to $\alpha$ for the states higher in the energy.  

\begin{figure}[t]
\begin{center}
\includegraphics*[width=0.8\columnwidth]{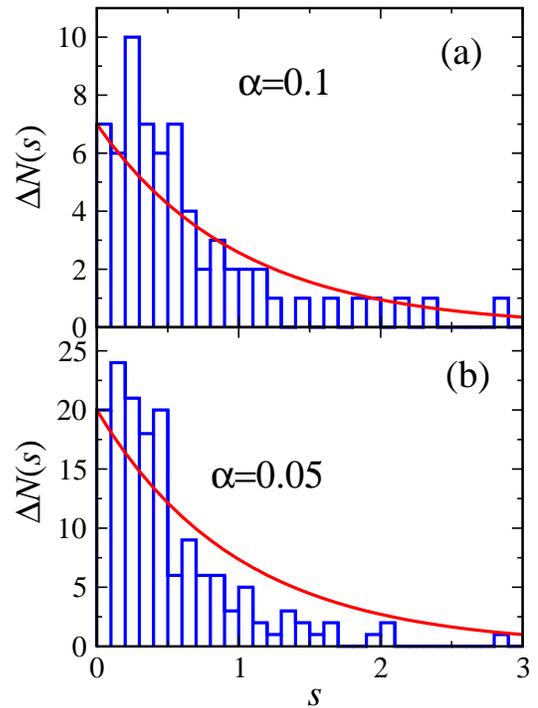}
\end{center}
\caption{Histogram for eigenenergies statistics in the
interval $\left( -2\protect\alpha ^{2},-0.51\protect\alpha ^{2}\right),$
where the effect of spin-orbit coupling is expected to be strong: (a) $\alpha =0.1$
and (b) $\alpha =0.05.$ Solid red lines correspond to the Poissonian statistics $\sim\exp(-s).$
Note that both panels demonstrate a very similar behavior. 
In the panel (a) the total energy interval contains 73 energy levels, and in the panel (b) it includes 174 levels. 
}
\label{fig:statistics}
\end{figure}

To illustrate the relation to possible emergence of chaos, we analyze the statistics of the
spectrum in a narrow interval strongly influenced by the SOC as 
presented in Fig. \ref{fig:statistics}. {One can clearly see the distinct deviation from the Poisson distribution.
Even though not strong enough to suggest the emergence of quantum chaos, the level distribution shows that
the SOC affects the regular character of the motion. In Fig.~\ref{fig:statistics} we show the histograms 
for the two values of SOC, $\alpha = 0.05$ and $\alpha=0.1$, with the $174$ and $73$ energy levels 
taken from the energy interval $(-2\alpha^2, -0.51 \alpha^2)$, respectively.  
}
Similarly to Refs. [\onlinecite{Marchukov},\onlinecite{Marchukov2}], here we remove the
double degeneracy related to the time-reversal symmetry preserved by the spin-orbit coupling. 
{One can see that for stronger SOC the distribution becomes wider which means stronger energy levels repulsion.}
To quantify the effect of the level repulsion, we compare the histogram in Fig.\ref{fig:statistics} with that of 2D hydrogen atom
at $\alpha=0$, where states are highly degenerate. Taking into account that for large $n$ the interlevel distance 
$E_{n+1}-E_{n}=1/n^{3}+O\left(1/n^{5}\right)$ and 
choosing a single histogram interval $s-\Delta _{s}/2<\left(E_{n+1}-E_{n}\right)/\epsilon<s+\Delta_{s}/2,$
where $\epsilon$ is the mean distance between the adjacent levels and $s$ is the corresponding energy bin size, 
yields the boundaries for the states numbers $n_{\max}=\epsilon ^{-1/3}\left( s-\Delta
_{s}/2\right)^{-1/3},$ and $n_{\min}=\epsilon^{-1/3}\left(s+\Delta_{s}/2\right)^{-1/3}.$ 
The number of the states in the interval $\left(n_{\min},n_{\max }\right)$, $\Delta N$ 
given by $\Delta N=n_{\max }^{2}-n_{\min }^{2}$
yields $\Delta N=2\epsilon ^{-2/3}\Delta _{s}/3s^{5/3}$ and we obtain 
$\Delta N\sim \Delta _{s}s^{-5/3}$ rapidly increasing at small $s$. This is qualitatively
different from the results  shown in Fig. \ref{fig:statistics}, clearly demonstrating the level repulsion caused by the SOC,
which is the necessary condition of the quantum chaos. 

In general, three types of statistics are expected to describe systems in terms of the distances 
between the adjacent energy levels. 
Poissonian statistics $P_{P}(s)=\exp(-s)$ is expected in the
absence of chaos. Two other statistics such as the Gaussian Orthogonal
Ensemble $P_{\rm GOE}(s)=\pi/2\times s\exp(-\pi s^{2}/4)$ and the Gaussian Unitary
Ensemble $P_{\rm GUE}(s)=32/\pi^2\times s^{2}\exp(-4s^{2}/\pi)$ (observed in Ref. [\onlinecite{Assmann2016}]),
are expected for different quantum chaos realizations. For these two, the level repulsion should be sufficiently 
strong to yield $P_{\rm GOE}(0)=P_{\rm GUE}(0)=0.$
Although the distribution in Fig. \ref{fig:statistics}, being a result of the level repulsion,
clearly deviates from the Poisson distribution, this repulsion is not sufficiently strong \cite{Silva2015}  to suppress 
it at $s\ll1$.

Note that Refs. [\onlinecite{Marchukov},\onlinecite{Marchukov2}] examined in detail the spectral properties
of another 2D quantum system such as anisotropic harmonic oscillator with SOC 
in terms of the quantum chaos ensembles. However, the results for the harmonic and Coulomb 
potentials cannot be compared directly since their eigenstates  are qualitatively different 
in terms of spectrum and wave functions.

\begin{figure}[t]
\begin{center}
\includegraphics*[width=0.8\columnwidth]{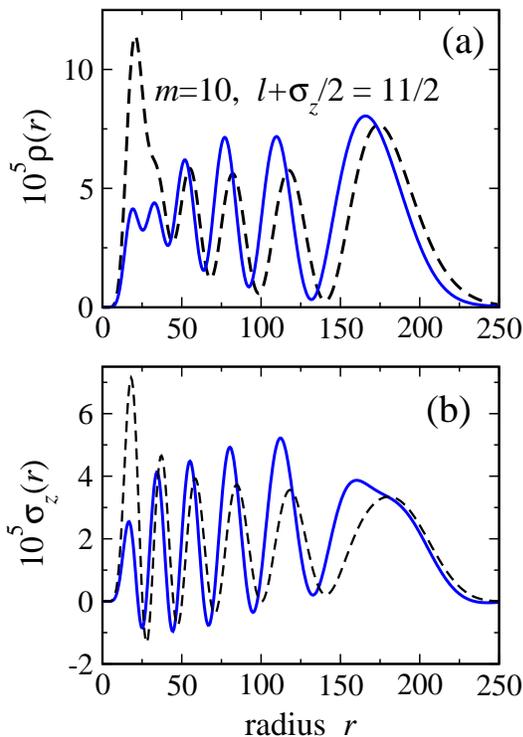}
\end{center}
\caption{Charge (a) and spin density (b) distributions for 
given eigenstate at $\alpha=0.1$. For comparison, we present by dashed lines the densities corresponding 
to correct zeroth-approximation functions, that is $\left(R_{11,5}^{2}+R_{11,6}^{2}\right)/2$
and $\left(R_{11,5}^{2}-R_{11,6}^{2}\right)/2$ for the density and for the $\sigma_{z}-$density, respectively.
These functions describe densities in the $\alpha\to 0^+$ limit. }
\label{fig:densities}
\end{figure}

The inclusion of SOC term into the 2D exciton Hamiltonian affects not only 
the spectral properties of the system but observables as well.
Consider local observables such as charge and spin densities
\begin{equation}
\sigma_{i,l}^{[m]}\left( r,\varphi \right) =\left( {\bm\psi }_{l}^{[m]}\left(
r,\varphi \right) \right) ^{\dag }\sigma _{i}{\bm\psi }_{l}^{[m]}\left(
r,\varphi \right).
\end{equation}%
The corresponding integral quantities read
\begin{equation}
\left\langle \sigma _{i}\right\rangle_{l}^{[m]} =\int \sigma _{i,l}^{[m]}\left( r,\varphi
\right) d^{2}r.
\end{equation}%
Here, the charge density is characterized by the identity matrix $\sigma _0$.
Note that $\langle \sigma_{x}\rangle_{l}^{[m]} =\langle\sigma_{y}\rangle_{l}^{[m]}=0$
due to orthogonality of the wave functions
corresponding to different values of $l$ contributing to the spinors. In
addition, the in-plane components of the spin have the symmetry 
$\sigma_{x,l}^{[m]}\left( r,\varphi \right) =\sigma_{l}^{[m]}(r)\cos \varphi $ and $\sigma
_{y,l}^{[m]}\left( r,\varphi \right) =\sigma_{l}^{[m]}(r)\sin \varphi ,$ respectively. Note 
that for Eq. \eqref{psispinor}: $\left\langle \sigma _{z}\right\rangle
_{l}^{[m]}=1-2\sum_{n}\left\vert c_{l,n\downarrow }^{[m]}\right\vert ^{2}<1.$

To illustrate the effect of the SOC on the 
eigenfunctions, we take a single eigenstate in the semiclassical domain and
show corresponding charge and spin densities in 
Fig. \ref{fig:densities}. As one can see, the 
behavior of the densities for finite $\alpha$ and those in the 
limit $\alpha\to \ 0$ are noticeably different. 

\begin{figure}[t]
\begin{center}
\includegraphics*[width=0.8\columnwidth]{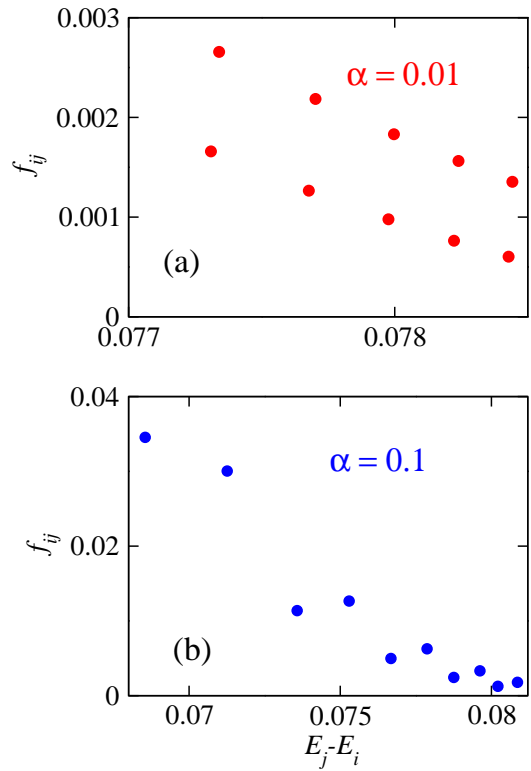}
\end{center}
\caption{Oscillator strengths for selected transitions with $l_{i}=2,$ $l_{j}=3$ 
for two different values of spin-orbit coupling, indicated in the panels.}
\label{fig:strength}
\end{figure}

Related quantity, which can be experimentally measured, is the oscillator
strength of the optical transition $f_{ij}=\left( E_{j}-E_{i}\right)
\left|x_{ij}\right|^{2},$ where $x_{ij}$ is the matrix element of the coordinate 
$x=r\cos \varphi \equiv r\left( e^{i\varphi }+e^{-i\varphi }\right) /2$ for transitions 
satisfying the selection rules: $l_j=l_i\pm\,1$.  We obtain for the transition $l_j=l_i+1$  
\begin{eqnarray}
&&f_{ij}=\sum_{n_{1}=n+1}\sum_{n=l+1}c_{l+1,n_{1}\uparrow
}^{[m_{1}]}c_{l,n\uparrow }^{[m]} r_{l \to l+1, n\to n_1} +  \nonumber \\
&&\sum_{n_{1}=n+1}\sum_{n=l+2}c_{l+1,n_{1}\downarrow
}^{[m_{1}]}c_{l,n\downarrow }^{[m]}
{r}_{l+1\rightarrow l+2,n\rightarrow n_{1}}, \label{me1} \\
&&{r}_{l+1\rightarrow l+2,n\rightarrow n_{1}} = \int
r^2R_{n_{1},l+2}(r)R_{n,l+1}(r)dr. \label{me2}
\end{eqnarray}%
Matrix elements \eqref{me2} can be calculated semiclassically similarly to Eq.
(\ref{GRint}). The results of numerical calculations are presented in Fig. \ref{fig:strength}.
At $\alpha=0.01$ the transitions demonstrate clear doublet structure. Although for
$\alpha=0.1$ the general structure of the spectrum, shown here up to the endpoint near the $-\alpha^{2}/2,$  
and the oscillator strengths distribution becomes less regular, the jigsaw-like pattern described in Ref. [\onlinecite{Gutzwiller}] 
does not appear here.

To make connections to the experiments, we consider GaAs-based
2D structures that are excellently suited for studies of excitonic spectra.\cite{High,Vishnevsky} In this case the characteristic speed 
$\sqrt{\left\vert E_{0}\right\vert /\mu }$ of the electron in the ground
state is of the order of $e^2/\kappa \sim 10^7$ cm/s. Also, the energy 
of the exciton ground state is of the order of -10 meV. Since the typical structure-dependent
values of $\alpha /\hbar $ are of the order 10$^{6}$ cm$^{-1}$, the
dimensionless $\alpha $ here is  about 0.1. Taking into account
that the velocity $\sqrt{\left\vert E_{n}\right\vert /\mu }$ decreases as $%
e^{2}/n\kappa$, we conclude that the states prone to chaos are
located at $n\sim 10$, corresponding to the present analysis.

{We are now in a position to discuss the effects 
related to the Berry curvature.\cite{zhou15} 
The common 2D systems in this case are monolayers of transition 
metals dichalcogenides like MX$_2$ (where M=Mo, W and X=S, Se) as they were 
used for estimations in Ref. [\onlinecite{zhou15}]. It has been shown 
experimentally \cite{nat14} that exciton binding energy, e.g., a in single layer 
WS$_2$ is close to 0.7 eV, which is extremely large as compared 
to the typical values of 10 meV for conventional semiconductors. In Ref. [\onlinecite{zhou15}], the 
role of SOC $\alpha$ is played by the quantity $\omega\nabla V$, where $\omega$ 
is the Berry curvature and $V({\bf r})$ is the screened Coulomb interaction. 
Denoting $\alpha_0\equiv\omega {\bar V}/a_0$ (where $a_0$ is the exciton Bohr 
radius and ${\bar V}$ is the average potential which 
could be around half of the exciton binding energy), we estimate 
this quantity to be $\alpha_0 \approx 0.26 $ eV$\rm{\AA}$ for the 
typical values $\omega \approx 15$ ${\rm\AA}^2$ and $a_0 \approx 20$ $\rm{\AA}.$\cite{nat14,zhou15} 
Note that this effective SOC constant $\alpha_0$ turns out to be 
less than that, e.g., for semiconducting perovskite 
layers $\alpha \approx 1$ eV$\rm{\AA}.$\cite{mypccp2018,r51} 
However, the effect of the term $\omega\nabla V$ on the spectra of Rydberg excitons
in conventional semiconductor heterostructures can be strongly different due to the
material dependence of the Berry $\omega-$parameter and the high spread of the $V-$potential. 
In general, the above estimations are vulnerable to many factors and primarily 
to the system and samples type. We postpone the discussion of the Berry curvature influence 
on chaos in 2D Rydberg excitons for future research. }

\section{Conclusions and outlook}

We demonstrated that spin-orbit coupling induces 
effective level repulsion resulting in a non-singular distribution of the distances between the 
adjacent energy levels and suggesting a possible crossover to a quantum chaotic regime. Although 
we did not observe the inherent in strong quantum chaos \cite{Gutzwiller} qualitative deviation 
of the levels statistics from the Poissonian one, the observed statistics can be a precursor to
emergence of the quantum chaos-related Gaussian Orthogonal or the Unitary ensembles. 
A possible reason for the observed distribution of the levels
is that the density of states  here is infinite as the energy approaches $-\alpha^2/2,$ 
and the level repulsion has to counteract this strong divergence. 
This situation is opposite to more conventional quantum chaotic systems such as trapped cold atoms 
or those with the orbital motion quantized by a magnetic field, where such a divergence does not occur. 

In particular, this result implies that quantization "suppresses" the manifestations of the 
classical chaos emerging in two-dimensional excitons due to the Rashba 
spin-orbit interaction \cite{mypccp2018}, meaning that the quantization makes the system more "ordered" than its  
classical counterpart. 
Thus, the quantization of a system demonstrating a classical chaos, may, in general, 
not generate as vivid chaos manifestations, as observed for its classical motion; and this is 
one of the main messages of this paper. To trace the origin of this difference, the semiclassical
Wentzel-Kramers-Brillouin-like analysis \cite{land3} of the quantum problem might 
be useful for the future studies. 

To get more insight into the possible quantum chaotic features of this system, 
the experimental studies of two-dimensional semiconductor structures with spin-orbit 
coupled Rydberg excitons are highly desirable. These studies can shed 
light on the corresponding energy levels statistics. We also note that the studies of quasistationary exciton states 
with the energies above the $-\alpha^2/2$ threshold  would be of interest 
as they can demonstrate other types of statistics in the real part of the eigenenergies.

\begin{acknowledgments}
E.Y.S. acknowledges the support by the Spanish Ministry of Economy, 
Industry, and Competitiveness and the European Regional 
Development Fund FEDER through Grant No. FIS2015-67161-P (MINECO/FEDER), 
and Grupos Consolidados UPV/EHU del Gobierno Vasco (IT-986-16).
N.T.Z. acknowledges support from the Aarhus University Research Foundation 
under the JCS Junior Fellowship and the Carlsberg Foundation through a 
Carlsberg Distinguished Associate Professorship grant.
\end{acknowledgments}

\end{document}